\documentclass[conference]{IEEEtran}
\IEEEoverridecommandlockouts

\usepackage{cite}
\usepackage{amsmath,amssymb,amsfonts}
\usepackage{algorithmic}
\usepackage{graphicx}
\usepackage{textcomp}
\usepackage{booktabs}

\usepackage{xcolor}
\usepackage[colorlinks=false,      
            linkcolor=black,        
            urlcolor=blue,        
            citecolor=blue        
           ]{hyperref}
\usepackage[ruled,linesnumbered,vlined]{algorithm2e}
\usepackage{amsmath}
\SetKwInput{KwIn}{Input}
\SetKwInput{KwOut}{Output}
\SetKwComment{SideComment}{$\blacktriangleright$\ }{}

\def\BibTeX{{\rm B\kern-.05em{\sc i\kern-.025em b}\kern-.08em
    T\kern-.1667em\lower.7ex\hbox{E}\kern-.125emX}}
\begin{document}

\title{AnatoMaskGAN: GNN-Driven Slice Feature Fusion and Noise Augmentation for Medical Semantic Image Synthesis\\
}


\author{
    Zonglin Wu\textsuperscript{1}\thanks{Zonglin Wu: noheadwuzonglin@gmail.com} , Yule Xue\textsuperscript{1}, Qianxiang Hu\textsuperscript{1}, Yaoyao Feng\textsuperscript{1}, Yuqi Ma\textsuperscript{1}, Shanxiong Chen\textsuperscript{1}* \\
    \textsuperscript{1}College of Computer and Information Science, Southwest University,
}

\maketitle               

\begin{abstract}
Medical semantic-mask synthesis boosts data augmentation and analysis, yet most GAN-based approaches still produce one-to-one images and lack spatial consistency in complex scans. To address this, we propose AnatoMaskGAN, a novel synthesis framework that embeds slice-related spatial features to precisely aggregate inter-slice contextual dependencies, introduces diverse image-augmentation strategies, and optimizes deep feature learning to improve performance on complex medical images. Specifically, we design a GNN-based strongly correlated slice-feature fusion module to model spatial relationships between slices and integrate contextual information from neighboring slices, thereby capturing anatomical details more comprehensively; we introduce a three-dimensional spatial noise-injection strategy that weights and fuses spatial features with noise to enhance modeling of structural diversity; and we incorporate a grayscale-texture classifier to optimize grayscale distribution and texture representation during generation. Extensive experiments on the public L2R-OASIS and L2R-Abdomen CT datasets show that AnatoMaskGAN raises PSNR on L2R-OASIS to 26.50 dB (0.43 dB higher than the current state of the art) and achieves an SSIM of 0.8602  on L2R-Abdomen CT—a 0.48 percentage-point gain over the best model, demonstrating its superiority in reconstruction accuracy and perceptual quality. Ablation studies that successively remove the slice-feature fusion module, spatial 3-D noise-injection strategy, and grayscale-texture classifier reveal that each component contributes significantly to PSNR, SSIM, and LPIPS, further confirming the independent value of each core design in enhancing reconstruction accuracy and perceptual quality.
\end{abstract}


\section{Introduction}
Medical imaging underpins diagnosis, monitoring, treatment planning, and education by revealing internal anatomy and pathology \cite{puttagunta2021medical}. Deep learning has boosted segmentation, classification, and computer-aided diagnosis, yet it hinges on large, expertly annotated datasets. Equipment costs, privacy rules, and labor-intensive labeling keep data supply far below demand, constraining model generalization and accuracy—a major bottleneck for medical AI \cite{zhou2021review}. Thus, developing effective data-expansion and augmentation methods under limited-data conditions has become a pressing research focus.

To tackle these challenges, researchers increasingly use semantic-mask–conditioned medical image generation \cite{zhou2021review}. Detailed masks of anatomy and lesions steer models to create realistic, semantically correct images, enriching data diversity and authenticity while sharply cutting acquisition and annotation costs, thereby supporting data augmentation and computer-aided diagnosis\cite{shin2018medical,dumont2021overcoming}.

In semantic-mask synthesis, Generative Adversarial Networks (GANs)\cite{goodfellow2014generativeadversarialnetworks} have become a research focus because of their unique one-to-many inference pattern, highly controllable latent space, and ability to generate high-resolution images from small samples. Specifically, a GAN can produce multiple stylistically different medical images in a single forward pass lasting only milliseconds, greatly improving data-expansion efficiency. The latent-space properties support fine control and linear interpolation of semantic masks, enabling precise reconstruction of local anatomical details and lesion characteristics. Moreover, GANs possess strong feature-extraction and representation capabilities, allowing them to capture sub-millimeter structural details and lesion features even in small-sample settings, showing clear potential over other generation techniques. Applying GANs to medical semantic-mask synthesis therefore promises revolutionary advances in healthcare.

Nevertheless, most existing GAN-based methods still use a one-to-one mapping—each semantic mask yields a single 2D image slice (i.e., a cross-sectional scan of a 3D volume). When applied to high-dimensional, multi-slice medical data, this simple mapping cannot enforce spatial consistency between adjacent slices or leverage anatomical priors (pre-learned knowledge of typical organ shapes and relationships). As a result, generated images often suffer from poor structural coherence, imprecise boundaries, and missing fine details. In particular, without effective contextual fusion across slices and explicit anatomical constraints, tissue regions can become discontinuous, details may vanish, and textures appear inconsistent. Moreover, traditional GANs typically inject only a single noise vector, limiting their ability to capture variability in structure and texture—thus hindering both generalization and clinical usefulness.

To systematically address these problems and improve semantic-mask synthesis for complex medical images, we propose a novel framework—AnatoMaskGAN. First, we design an innovative graph-neural-network (GNN) slice-feature fusion module, which constructs a spatial adjacency graph between slices and uses a GNN to efficiently aggregate contextual information from neighboring slices. This builds rich spatial dependencies, more accurately capturing continuity of anatomical structures and lesion regions, and markedly enhancing spatial coherence and anatomical consistency in the generated images. Second, we introduce a spatial three-dimensional noise-injection strategy that fuses weighted random noise in the spatial domain, strengthening the model’s ability to represent diverse anatomical structures and tissue textures and improving generalization. Finally, to optimize visual-perception quality, we add a grayscale-texture classifier that explicitly constrains the learning of grayscale distributions and texture features, bringing generated images closer to real medical images in grayscale detail and textural realism.

On L2R-OASIS (brain MRI), AnatoMaskGAN raises PSNR to 26.50 dB—0.43 dB above the state-of-the-art—while on L2R-Abdomen CT it lifts SSIM to 0.8602, a 0.48 \% gain. Ablations show that slice-feature fusion, 3-D noise injection, and the gray-level/texture classifier each add essential, complementary boosts across PSNR, SSIM, and LPIPS.

In summary, our contributions are threefold:
\begin{itemize}
  \item \textbf{GNN-based slice-feature fusion:} deep modeling of spatial context significantly improves anatomical continuity.
  \item \textbf{3-D spatial noise injection:} enhances generalization to complex structures and diverse textures.
  \item \textbf{Grayscale-texture classifier:} boosts photorealism and fine-grained texture fidelity in the synthesized images.
\end{itemize}

\section{RELATED WORK}
In recent years, the explosive growth of deep-learning techniques has propelled remarkable progress in medical-image analysis. Yet the field’s unique constraints—strict data privacy, the difficulty and expense of expert annotation, and high acquisition costs—mean that medical-imaging datasets remain limited in both scale and diversity, curbing the real-world clinical performance of deep-learning models. To mitigate data scarcity, researchers have increasingly turned to generative-adversarial-network (GAN)–based augmentation. After Goodfellow et al. introduced the GAN framework \cite{goodfellow2014generativeadversarialnetworks}, it was rapidly adopted for medical-image data expansion, including cross-modal image translation \cite{armanious2020medgan,yu2019ea,salehjahromi2024synthetic} and intra-modal augmentation \cite{jiang2020covid,10.1007/978-3-031-53767-7_21,konz2024anatomically}, showing encouraging gains in data enrichment and diagnostic support.

Early GANs used basic adversarial losses and weak structural constraints, so their outputs often had blurred boundaries and anatomically implausible distortions. Conditional GANs \cite{mirza2014conditional} addressed this by conditioning on semantic masks—pix2pix \cite{isola2017image}, for instance, guides the generator with a segmentation map—while SPADE \cite{park2019semantic} improved granularity by injecting map-derived modulation parameters into each normalization layer, giving every semantic region its own activations. Although SPADE now underpins many models and excels at both natural- and medical-image synthesis, its locally focused modulation still struggles to capture long-range dependencies, leaving complex layouts prone to semantic gaps and fuzzy edges.

Building on SPADE, Hao Tang et al. proposed the Dual-Attention GAN (DAGAN) \cite{tang2020dual}, introducing spatial- and channel-attention modules to capture long-range dependencies and region-specific features, thereby producing higher-quality, semantically consistent images. Although DAGAN preserves global structures in complex semantic maps, it remains highly dependent on the geometric layout of the mask and still models appearance correlations between semantic regions only weakly. To address this, Yi Wang et al. presented SCGAN \cite{wang2021image}, which adds convolutional normalization to obtain features with both semantic correlation and spatial variability, enhancing coherence and diversity across semantic regions. SCGAN boosts generation quality while retaining controllability, but its reliance on multiple perceptual losses leads to complex training and high computational cost.

Seeking a more streamlined approach, Vadim Sushko et al. introduced OASIS \cite{sushko2020you}, redesigning the discriminator as a semantic-segmentation network that takes the label map itself as ground truth, allowing it to pinpoint semantic deviations in generated images and provide the generator with finer spatial feedback. OASIS also combines local- and global-sampling strategies with a 3-D noise tensor to enable multi-modal appearance control for single semantic regions, achieving strong results in natural-image tasks. However, it still falls short in rendering fine-grained structures. Zhengyao Lv et al. therefore proposed SAFM \cite{lv2022semantic}, incorporating a Shape-aware Position Descriptor (SPD) and a Semantic–Shape Adaptive Feature Modulation (SAFM) module. SPD captures pixel-level shape structures and latent part layouts, while SAFM fuses semantic information with shape-position cues to modulate features adaptively, markedly improving boundary realism and structural plausibility.

To tackle these challenges we propose AnatoMaskGAN, which retains the SPADE backbone while integrating graph neural networks (GNNs)\cite{scarselli2008graph} to model cross-slice spatial features deeply, markedly improving the generator’s perception of anatomical continuity. We introduce a spatially weighted 3-D noise tensor to enrich texture diversity and region-specific appearance without disturbing global structure. Finally, a specialized grayscale-texture classifier discriminatively refines grayscale distribution and textural detail, enhancing radiological realism from a statistical perspective.
\begin{figure*}[htbp]
  \centering
  \includegraphics[width=\textwidth,
                   height=.35\textheight,
                   keepaspectratio]{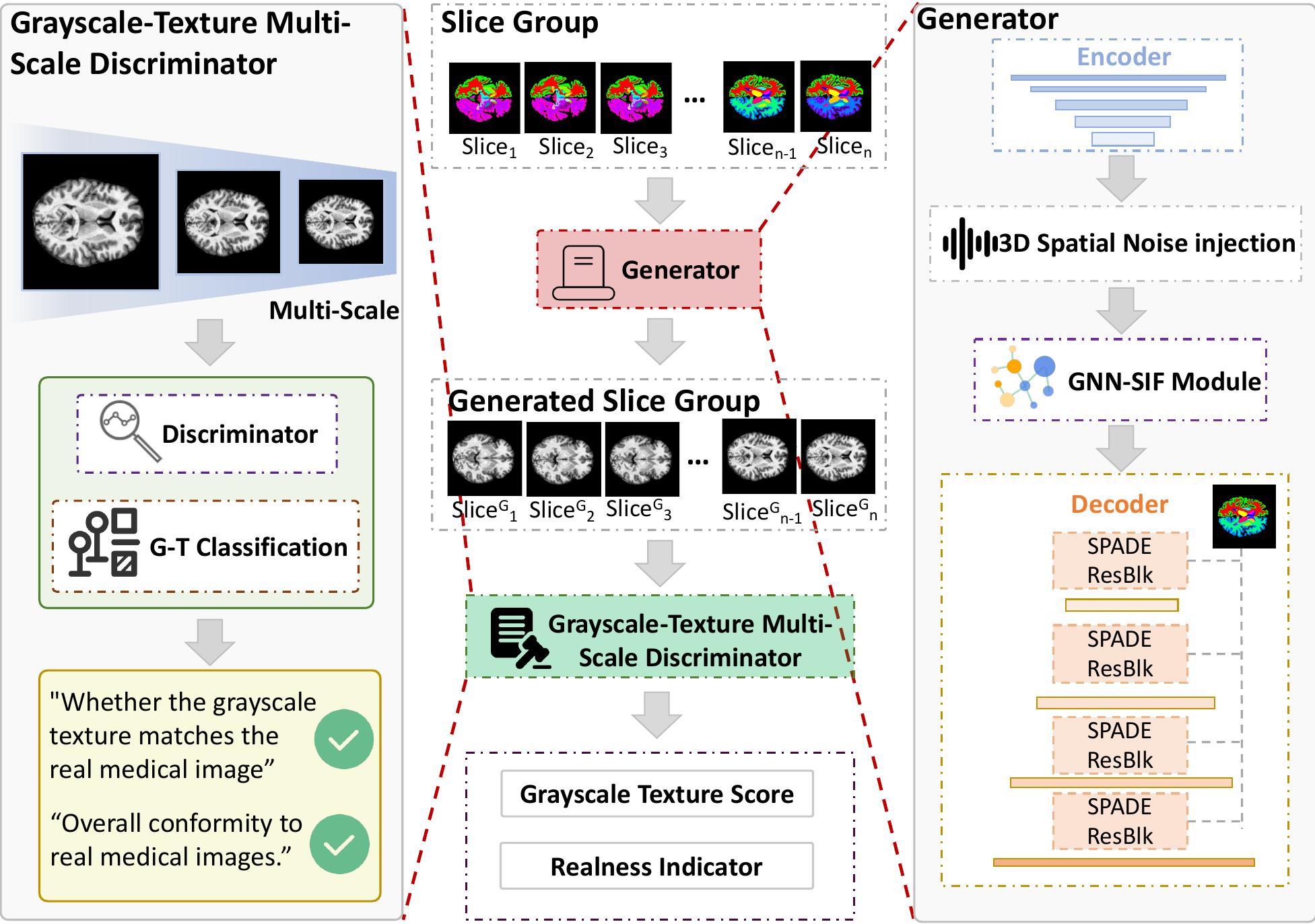}
  \caption{The overall architecture of our proposed method, AnatoMaskGAN.}
  \label{fig1}
\end{figure*}
\section{METHODS}
\paragraph{AnatoMaskGAN.} 
Figure~\ref{fig1} overviews our SPADE-based framework, which operates on a contiguous stack of $n$ semantic-mask slices $\{\mathrm{Slice}_i\}_{i=1}^{n}$ and comprises \emph{generation} and \emph{discrimination} stages.  
An encoder first extracts high-dimensional semantics from each slice.  A \textbf{3-D spatial noise-injection} module (3D-SNI) aligns Gaussian noise with the volume’s physical coordinates and injects it into the features, yielding fine yet inter-slice-consistent texture variation.  
To capture anatomical continuity, a \textbf{GNN slice-integration} module (GNN-SIF) treats slices as graph nodes and fuses their features according to edge weights encoding physical distance and semantic similarity.  The fused representation is decoded by SPADE-modulated deconvolutions to produce the grayscale volume $\{\mathrm{Slice}_i^{G}\}_{i=1}^{n}$.  

For realism, a multi-scale structural discriminator is paired with a \textbf{gray-texture classifier} (G-TC) that judges gray-level histograms and local texture vectors, giving the generator explicit feedback on intensity statistics and detail.  
The cooperation of 3D-SNI, GNN-SIF, and G-TC delivers state-of-the-art semantic controllability, structural coherence, texture diversity, and radiological fidelity.

\subsection{GNN-Based Strongly Interdependent Slice Feature Fusion Module(GNN-SIF)}\label{AA}
\begin{figure*}[htbp]
  \centering
  \includegraphics[width=\textwidth,
                   height=.29\textheight,
                   keepaspectratio]{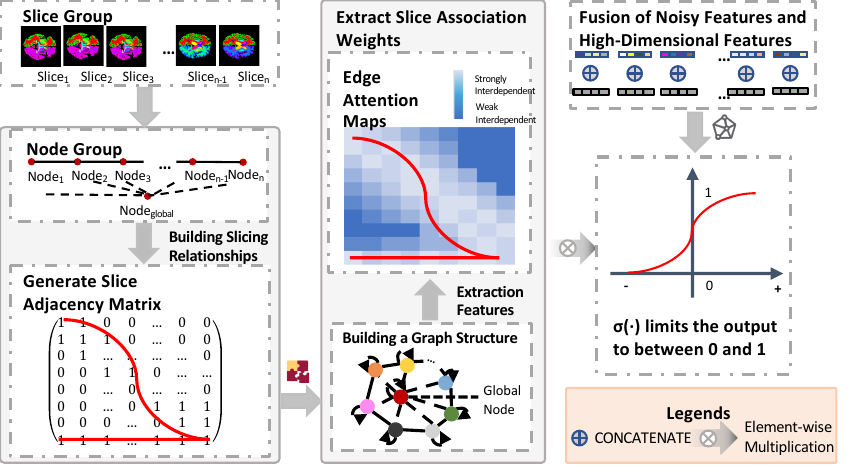}
  \caption{Specific implementation process of the GNN-SIF module.}
  \label{fig3}
\end{figure*}
Traditional one-to-one medical-image generation methods often overlook the rich spatial correlations between adjacent slices, resulting in structural discontinuities, missing anatomical details, and blurry boundaries. To break this bottleneck, we propose a \emph{Graph Neural Network} (GNN)-based \textbf{Strongly-Correlated Slice Feature Fusion} module (GNN-SIF) that explicitly models the complex dependencies among slices and thus enhances spatial coherence and anatomical consistency. As illustrated in Fig.~\ref{fig3} and detailed in Algorithm~\ref{alg:gnn-sif}, GNN-SIF first treats each consecutive slice as a spatial node and adds a global node to aggregate holistic context; edges are assigned according to spatial proximity so that directly adjacent slices are strongly correlated while distant slices are weakly or non-correlated. The resulting adjacency matrix $\tilde{\mathbf A}$ is projected into a feature space to form edge-attention maps, which are then passed through a non-linear mapping to produce an attention distribution in $[0,1]$. Finally, this distribution is fused element-wise with high-dimensional feature maps to obtain the modulated output, formally expressed as
\begin{equation}
\begin{split}
x'_{i,j,k}=x_{i,j,k}\,\sigma\Bigl(&
  \sum_{n=1}^{N_{1}}
    W_{(i,j,k),n}\,
    \delta\Bigl(
      \sum_{m=1}^{N^{2}}
        U_{n,m}\,
        \mathrm{vec}(\tilde{\mathbf A})_{m}
      + b_{n}
    \Bigr) \\
  &\; + c_{i,j,k}
\Bigr)
\end{split}
\label{eq:gnn-sif}
\end{equation}
where $\tilde{\mathbf A}$ is the slice-graph adjacency matrix, $\mathrm{vec}(\tilde{\mathbf A})\!\in\!\mathbb{R}^{N^{2}}$ its vectorised form, $i\!\in\!\{1,\dots,C\}$, $j\!\in\!\{1,\dots,H\}$ and $k\!\in\!\{1,\dots,W\}$ index channels, height and width, respectively; $\delta(\cdot)$ and $\sigma(\cdot)$ denote an optional non-linear activation and the Sigmoid function; $U\!\in\!\mathbb{R}^{N_{1}\times N^{2}}$, $b\!\in\!\mathbb{R}^{N_{1}}$, $W\!\in\!\mathbb{R}^{(C H W)\times N_{1}}$ and $c\!\in\!\mathbb{R}^{C H W}$ are learnable parameters.
\begin{algorithm*}[t]   
\small                 
\caption{GNN-Based Strongly Interdependent Slice Feature Fusion (GNN-SIF)}
\label{alg:gnn-sif}
\KwIn{Medical image slices $X$, adjacency matrix $\tilde{A}$, learnable parameters $W,U,b,c$}
\KwOut{Fused feature map $F_{\mathrm{final}}$, feature map size $(C \times H \times W)$}

\BlankLine
\textbf{Step 1: Multi-scale Feature Aggregation}\;
$F_{\mathrm{fusion}} \gets X + Y$\SideComment{Add encoder and decoder features}
$F_1 \gets \mathrm{Cat}(F_{\mathrm{fusion}},
        \mathrm{DConv}(F_{\mathrm{fusion}}, r{=}2, C/2))$\SideComment{$F_1$ size: $H \times W \times C/2$}
$F_2 \gets \mathrm{Cat}(F_1,
        \mathrm{DConv}(F_1, r{=}4, C/4))$\SideComment{$F_2$ size: $H \times W \times C/4$}
$F_3 \gets \mathrm{Cat}(F_2,
        \mathrm{DConv}(F_2, r{=}8, C/8))$\SideComment{$F_3$ size: $H \times W \times C/8$}
$F_{\mathrm{mfa}} \gets \mathrm{Conv}(F_3, C)$\SideComment{$F_{\mathrm{mfa}}$ size: $H \times W \times C$}

\BlankLine
\textbf{Step 2: GNN-based Slice Feature Fusion}\;
$\tilde{A} \gets \mathrm{BuildAdjacencyMatrix}(X)$\SideComment{Build adjacency matrix}
$\text{nodes} \gets \mathrm{DefineNodes}(X)$\SideComment{Define nodes for each slice}
$F_{\mathrm{mfa\_nodes}} \gets \mathrm{GNNFeatureFusion}(F_{\mathrm{mfa}},\tilde{A})$\SideComment{Fuse features with GNN}

\BlankLine
\textbf{Step 3: Final Feature Adjustment}\;
$F_{\mathrm{adj}} \gets \mathrm{ApplyNonlinearActivation}(F_{\mathrm{mfa\_nodes}}, \delta)$\SideComment{Apply activation (e.g.\ ReLU)}
$F_{\mathrm{final}} \gets \mathrm{Conv}(F_{\mathrm{adj}}, 2C)$\SideComment{Final convolution}

\KwRet{$F_{\mathrm{final}}$}\;
\end{algorithm*}
\subsection{3D Spatial Noise Injection}
\begin{figure}[htbp]
  \centering
  \includegraphics[width=\columnwidth]{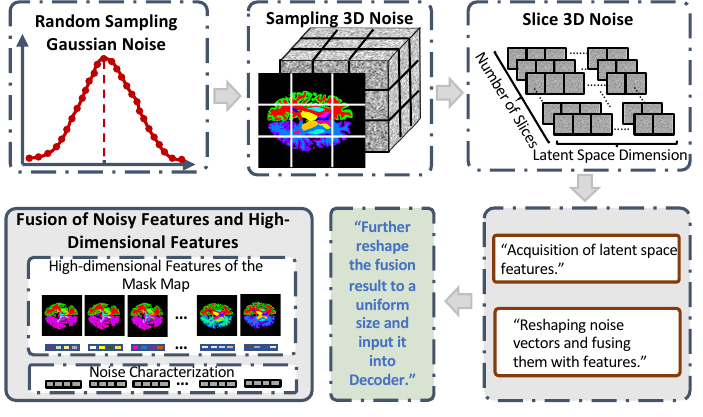}
  \caption{Spatial three-dimensional noise-injection mechanism.}
  \label{fig2}
\end{figure}

To mitigate the \emph{mode-collapse} phenomenon in medical-image generation—where a generator keeps producing visually similar outputs—we introduce a \textbf{spatial three-dimensional noise-injection (3D-SNI) strategy} for \textbf{AnatoMaskGAN}, as illustrated in Fig.~\ref{fig2}. Rather than adding independent random perturbations slice by slice, we first construct a noise volume $Z_{3}$ that matches the full three-dimensional scan. The noise is drawn from a standard Gaussian distribution and linearly mapped so that each voxel is precisely aligned with its three-dimensional coordinates. Owing to spatial continuity, adjacent voxels share similar characteristics, whereas voxels farther apart are only weakly correlated.
Next, $Z_{3}$ is partitioned along the $z$-axis, giving every semantic-mask slice its own noise chunk. Because the noise has already been smoothed in three-dimensional space, this design eliminates the “frame-jump” artifacts typical of conventional per-slice noise injection, thereby preserving textural continuity. Noise is injected through a residual modulation scheme:
\begin{equation}
H' = H + \bigl(Z_{3} \odot \sigma\!\bigl(F_{1}(Z_{3}) + W_{1}\bigr)\bigr) + \alpha\, R(Z_{3}),
\label{eq:3dsni}
\end{equation}
where $H$ is the current feature map, $Z_{3}$ the three-dimensional noise, $F_{1}(Z_{3})$ a nonlinear mapping of the noise, $W_{1}$ the transformation weights, $\alpha$ a weighting coefficient, and $R(Z_{3})$ the residual-noise component.

\subsection{Grayscale–texture joint classification (G-TC).}
\begin{figure}[htbp]
  \centering
  \includegraphics[width=\columnwidth]{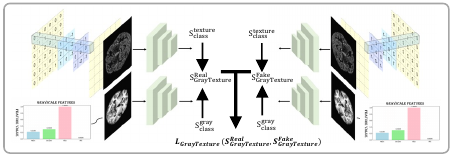}
  \caption{Gray-scale value and texture feature optimisation module: texture features are extracted from the original image and the generated image using the Sobel operator. The texture features and gray-scale image are fed into the classifier separately to obtain texture scores and gray-scale scores. The score loss is calculated using the scores of the two images.}
  \label{fig4}
\end{figure}
Although adversarial loss and feature-perceptual loss can preserve the \emph{macro-level} anatomy, generated medical images often exhibit a “plastic” appearance at the \emph{micro-level}: local textures lack the subtle undulations of real scans and boundaries appear soft or even blurred.
To address this long-standing difficulty, we insert a \textbf{grayscale–texture joint classification branch} (G-TC, Fig.~\ref{fig4}) alongside the discriminator, providing the generator with fine-grained feedback from grayscale statistics and gradient-based texture information.

\medskip
\noindent\textbf{Texture channel.}
We convolve the input image with Sobel kernels to obtain horizontal and vertical gradients,
$\nabla I_{i,x}$ and $\nabla I_{i,y}$, for each anatomical class~$i$.
The gradient magnitude is
$\lvert\nabla I\rvert=\sqrt{\nabla I_{x}^{2}+\nabla I_{y}^{2}}$.
Using the one-hot mask $\mathrm{mask}_{i}$ to restrict the magnitude to its region,
we define the intra-class texture score
\begin{equation}
\begin{aligned}
S_{\text{class}}^{\text{texture}}
 &= \sum_{k=1}^{2}\!
    \bigl(
      \sum_{i=1}^{N}\!
      \bigl(
        \int_{\Omega}
          (\nabla I_{i,x}\odot\nabla I_{i,y}+\epsilon)\,
          \mathrm{mask}_{i}\,d\mathbf{p}
\\
 &\hphantom{=\sum_{k=1}^{2}\!\bigl(\sum_{i=1}^{N}\!\bigl(}
      \bigr)^{\alpha_{k}}
    \bigr),
\end{aligned}
\label{eq:texture}
\end{equation}

where $\epsilon$ prevents division-by-zero, $N$ is the number of gradient directions,
$\alpha_{k}$ are texture weights, and $\Omega$ is the image domain.

\medskip
\noindent\textbf{Grayscale channel.}
For each class we mask the region and compute
mean $\mu_{j}$, standard deviation $\sigma_{j}$,
maximum $\mathrm{max}_{j}$, and minimum $\mathrm{min}_{j}$ at scale~$j$.
The grayscale score is
\begin{equation}
S_{\text{class}}^{\text{gray}}
    = \sum_{j=1}^{M}
      \left(
        \frac{\mu_{j}\,\sigma_{j}\,\mathrm{max}_{j}\,\mathrm{min}_{j}}
             {\sum \mathrm{mask}_{j}}
      \right)^{\beta_{k}},
\label{eq:gray}
\end{equation}
with $\beta_{k}$ weighting the grayscale features.
The G-TC branch linearly combines the weighted texture and grayscale
scores to produce the final class score $S_{\text{class}}$.

\medskip
\noindent\textbf{Overall objective.}
The generator is optimised with the composite loss
\begin{equation}
\begin{aligned}
\mathcal{L}_{G}
 &= \mathcal{L}_{\mathrm{GAN}}
    + \frac{\lambda_{\mathrm{feat}}}{N_{D}}
      \sum_{i=1}^{N_{D}}\sum_{j=1}^{L_{i}-1}
      \mathcal{L}_{\mathrm{feat}}^{(i,j)}
\\
 &\quad
    + \lambda_{\mathrm{VGG}}\,\mathcal{L}_{\mathrm{VGG}}
    + \lambda_{\mathrm{G\text{-}TC}}\,
      \mathcal{L}_{\mathrm{GrayTexture}},
\end{aligned}
\label{eq:loss}
\end{equation}

where $\mathcal{L}_{\mathrm{GAN}}$ is the adversarial loss,
$\mathcal{L}_{\mathrm{feat}}^{(i,j)}$ the feature loss from
discriminator~$i$ at layer~$j$,
$\mathcal{L}_{\mathrm{VGG}}$ the perceptual loss,
and $\mathcal{L}_{\mathrm{GrayTexture}}$ the loss from the proposed G-TC branch.
The hyper-parameters $\lambda_{\mathrm{feat}}$, $\lambda_{\mathrm{VGG}}$,
and $\lambda_{\mathrm{G\text{-}TC}}$ balance the respective terms;
$N_{D}$ is the number of discriminators and $L_{i}$ the number of layers
in discriminator~$i$.

\section{RESULTS}
\subsection{Datasets}
To verify the effectiveness and feasibility of the proposed method, we carried out experiments on the following two datasets.

\begin{figure}[htbp]
  \centering
  \includegraphics[width=\columnwidth]{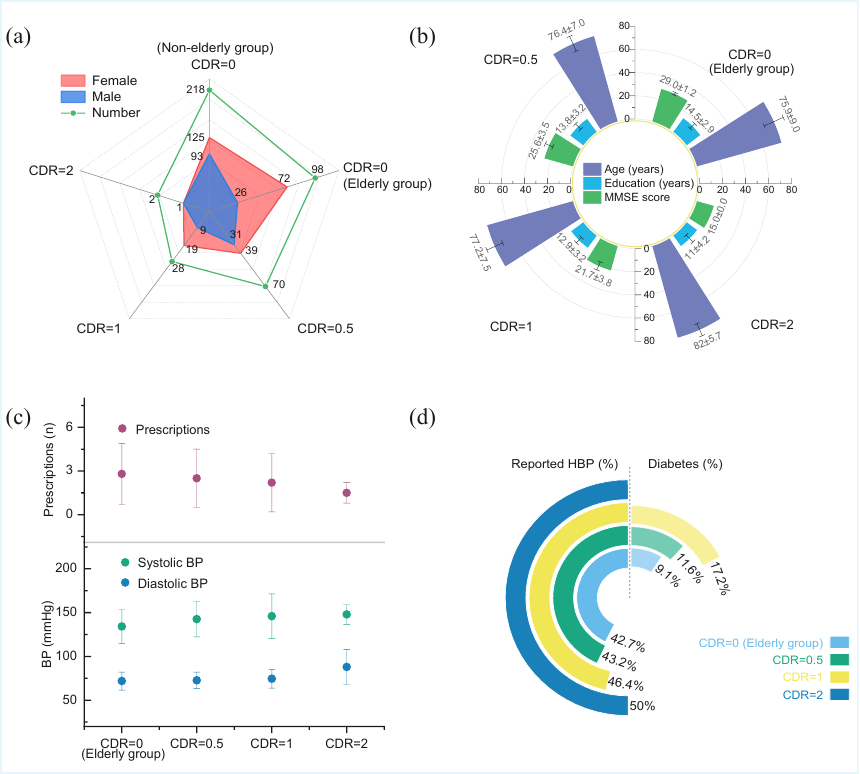}
  \caption{HBP: high blood pressure; CDR: Clinical Dementia Rating Scale (0=no dementia, 0.5/1/2/3=very mild/mild/moderate/severe dementia); As shown in Figure b, the MMSE scale score range from 30 points (best) to 0 points (worst); As shown in Figure c, Prescriptions (n) = number of prescription medications currently taking per subject; Systolic BP (mmHg): systolic blood pressure; Diastolic BP (mmHg): diastolic blood pressure; As shown in Figure d, Reported HBP (\%): prevalence of hypertension; Diabetes (\%): Prevalence of diabetes.}
  \label{fig7}
\end{figure}

L2R-OASIS dataset \cite{marcus2007open}: The L2R-OASIS dataset contains 416 three-dimensional T1-weighted MRI volumes, each annotated with 35 brain subregions and accompanied by corresponding ground-truth masks. The cross-sectional scans come from adults aged 18–96 years: 218 young and middle-aged subjects (18–59 years, all non-demented) and 198 older subjects (60–96 years, comprising 98 non-demented individuals and 100 patients with Alzheimer’s disease, AD). Descriptive statistics are reported as mean ± standard deviation, with ranges given in parentheses. Compared with cognitively normal elders, the demented group shows significantly lower MMSE scores and slightly fewer years of education. Detailed demographic and clinical characteristics of the 416 subjects are summarized in Fig. \ref{fig7}. All images were pre-processed with de-identification, motion correction, and spatial normalization. Each 3-D MRI volume was then sliced into 2-D images, and the corresponding semantic masks were generated.
Abdomen CT–CT dataset in L2R-Dataset (pre-2022)\cite{xu2016evaluation}: This dataset comprises 50 portal-venous-phase three-dimensional abdominal CT scans (30 for training and 20 for testing) with annotations for 13 abdominal organs, including the spleen, left and right kidneys, gallbladder, and others. All volumes were pre-processed through voxel-size harmonization, spatial normalization, and affine registration. Because annotations for the test set are not publicly available, only the training set was used in our experiments. 
\begin{figure*}[htbp]
  \centering
  \includegraphics[width=\textwidth,
                   height=.30\textheight,
                   keepaspectratio]{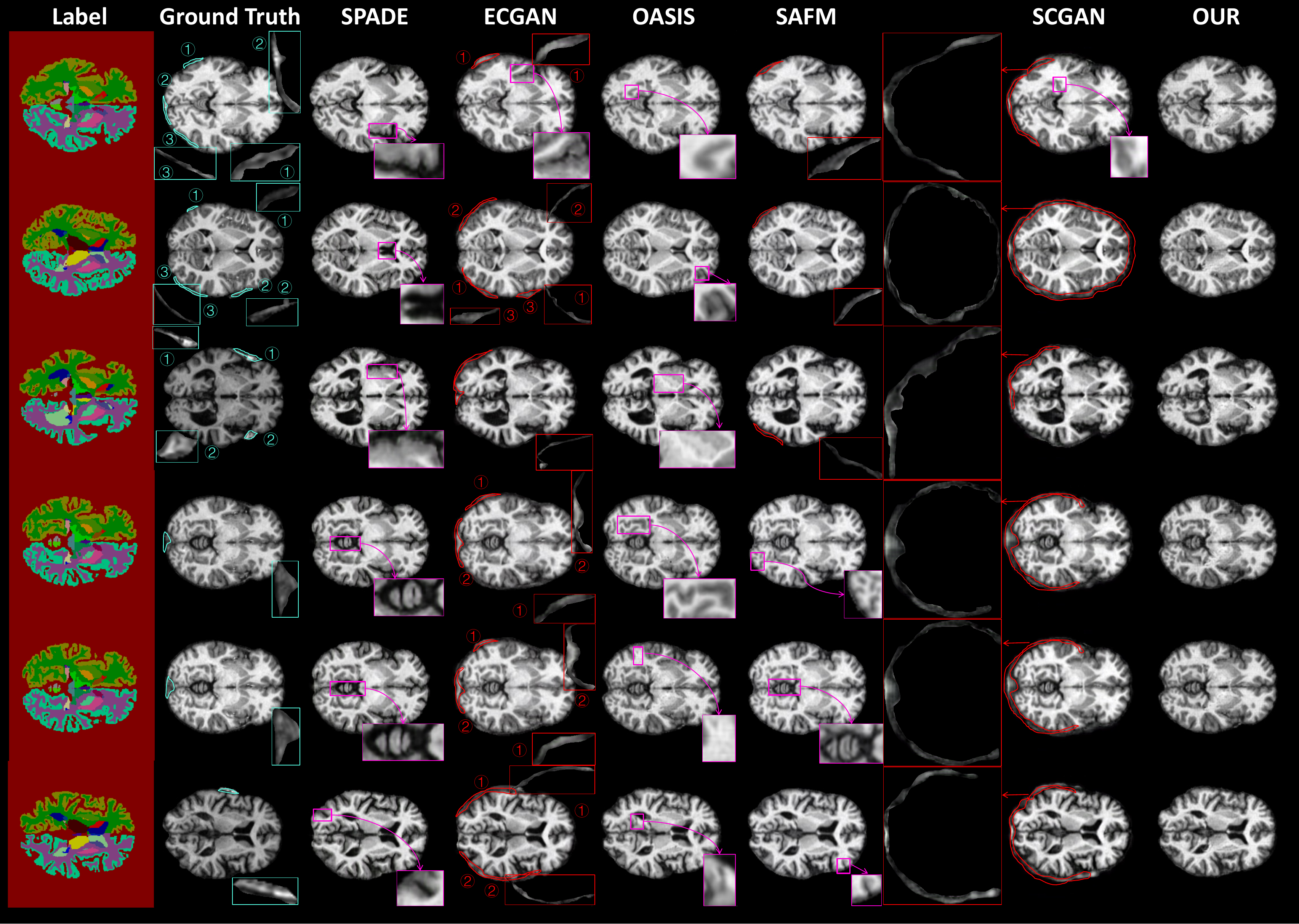}
  \caption{A collection of visualisations showing the generation effects of different models on the L2R-OASIS dataset.}
  \label{fig5}
\end{figure*}
\begin{figure*}[htbp]
  \centering
  \includegraphics[width=\textwidth,
                   height=.28\textheight,
                   keepaspectratio]{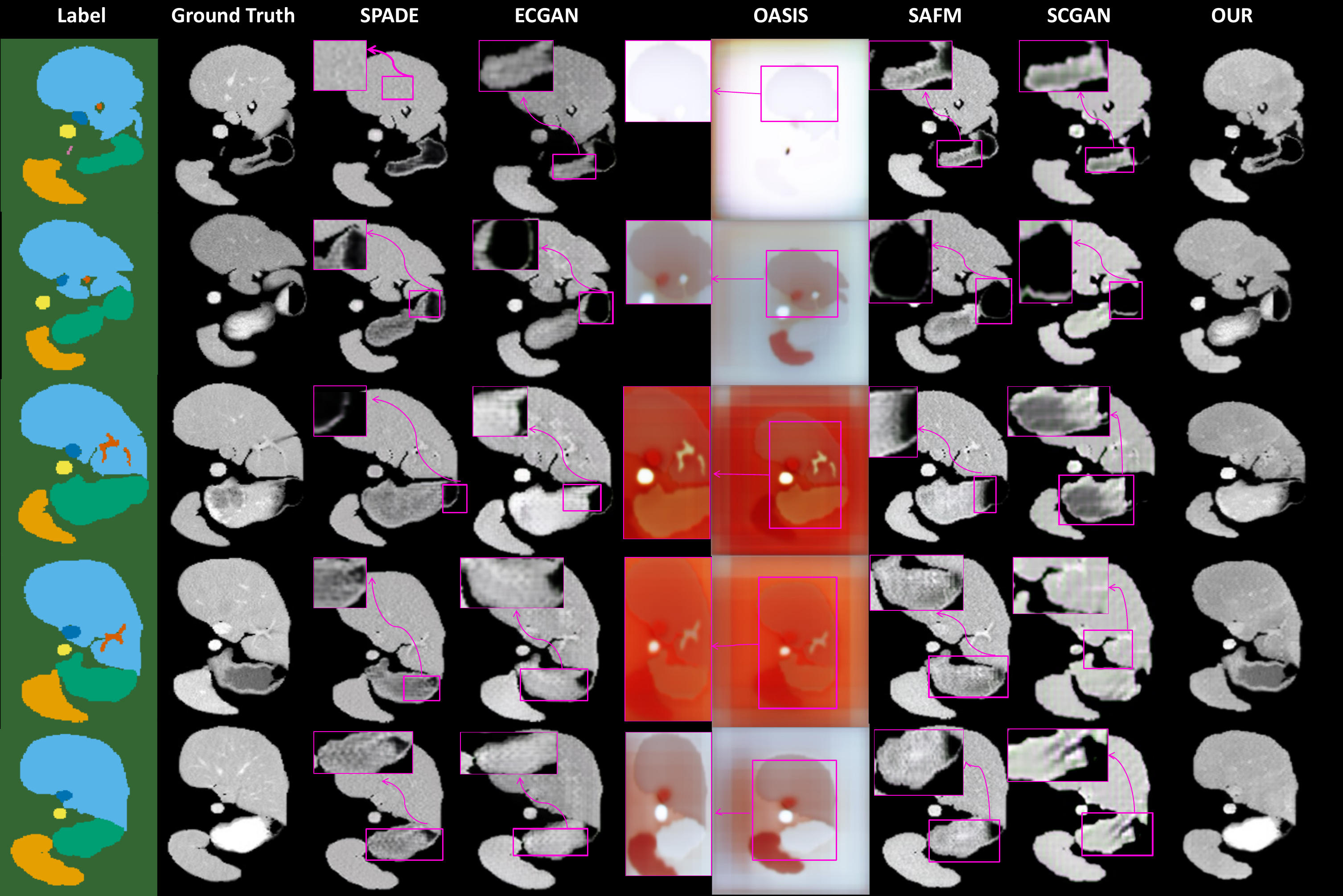}
  \caption{A collection of visualisations showing the generation effects of different models on the L2R-Abdomen CT dataset.}
  \label{fig6}
\end{figure*}
\subsection{Implementation Details}
All experiments are implemented using the PyTorch framework and trained on an NVIDIA A800 GPU. We use SPADE as the backbone network, with a total of 100 training iterations. The Adam optimizer with parameters \( \beta_1 = 0.1 \) and \( \beta_2 = 0.9 \) is used, and the learning rate is set to 0.0002 for optimization. Based on empirical results, we set \( \lambda_{\text{feat}} = 10 \), \( \lambda_{\text{vgg}} = 5 \), and \( \lambda_{\text{G-TC}} = 10 \). To further improve data diversity and model robustness, we apply data augmentation (such as rotation and flipping) to the slices of three datasets. The images are resized to 224x224 pixels, and normalization is performed to scale pixel values to the range [0, 1]. Finally, the dataset is split into training, validation, and test sets with a ratio of 70\%:15\%:15\%.

\begin{table*}[ht]
\centering
\caption{\textbf{Performance Comparison of Different Models on L2R-OASIS and L2R-Abdomen CT Datasets}}
\scriptsize
\resizebox{\textwidth}{!}{%
\begin{tabular}{lcccccc}
\toprule
\textbf{} & \multicolumn{3}{c}{\textbf{L2R-OASIS}} & \multicolumn{3}{c}{\textbf{L2R-Abdomen CT}} \\
\cmidrule(r){2-4} \cmidrule(l){5-7}
& PSNR (dB)↑ & SSIM↑ & LPIPS↓ & PSNR (dB)↑ & SSIM↑ & LPIPS↓ \\
\midrule
Spade (Baseline)    & 24.50 & 0.8871 & 0.1036 & 21.36 & 0.8550 & 0.0917 \\
SC-GAN              & 23.68 & 0.8579 & 0.1147 & 21.53 & 0.8553 & 0.1266 \\
EC-GAN              & 25.25 & 0.8909 & 0.0929 & 20.75 & 0.8484 & 0.0936 \\
D-GAN               & 25.61 & 0.8950 & 0.0908 & 21.07 & 0.8437 & 0.0913 \\
SAFM                & 25.84 & 0.9153 & 0.0784 & 22.35 & 0.8386 & 0.0896 \\
OASIS               & 26.07 & 0.9044 & 0.0922 & 20.45 & 0.8554 & 0.0991 \\
\textbf{G3‑Net (Ours)} & \textbf{26.50} & \textbf{0.9229} & \textbf{0.0559} & \textbf{21.98} & \textbf{0.8602} & \textbf{0.0807} \\
\bottomrule
\end{tabular}%
}
\label{tab:compare-results}
\end{table*}

\begin{table*}[ht]
\centering
\caption{\textbf{Ablation Study of AnatoMaskGAN on L2R-OASIS and L2R-Abdomen CT Datasets}}
\scriptsize
\resizebox{\textwidth}{!}{%
\begin{tabular}{lcccccc}
\toprule
\textbf{} & \multicolumn{3}{c}{\textbf{L2R-OASIS}} & \multicolumn{3}{c}{\textbf{L2R-Abdomen CT}} \\
\cmidrule(r){2-4} \cmidrule(l){5-7}
& PSNR (dB)↑ & SSIM↑ & LPIPS↓ & PSNR (dB)↑ & SSIM↑ & LPIPS↓ \\
\midrule
Spade (Baseline)             & 24.50 & 0.8871 & 0.1036 & 21.36 & 0.8550 & 0.0917 \\
Spade+GNN-SASE                  & 26.22 & 0.9115 & 0.0570 & 21.41 & 0.8560 & 0.1030 \\
Spade+GNN-SASE+3D Noise         & 26.42 & 0.9116 & 0.0567 & 21.43 & 0.8512 & 0.0919 \\
\textbf{Spade+GNN-SASE+3D Noise+G-TC} & \textbf{26.50} & \textbf{0.9229} & \textbf{0.0559} & \textbf{21.98} & \textbf{0.8602} & \textbf{0.0807} \\
\bottomrule
\end{tabular}%
}
\label{tab:ablation}
\end{table*}
\subsection{Evaluation Metrics}
To validate the effectiveness of our proposed method, we quantitatively evaluate the quality of the generated images from multiple dimensions. Specifically, we use three mainstream image quality evaluation metrics: Peak Signal-to-Noise Ratio (PSNR), Structural Similarity Index Measure (SSIM), and Learned Perceptual Image Patch Similarity (LPIPS). These three metrics measure the differences between the generated and real images from pixel-level, structural-level, and perceptual-level perspectives, respectively.

PSNR is a traditional and widely used image quality evaluation metric, which measures pixel-level errors between images. The formula for PSNR is:

\begin{equation}
\text{PSNR} = 10 \cdot \log_{10} \left( \frac{MAX^2}{\text{MSE}} \right) \quad
\end{equation}

where \( MAX \) is the maximum pixel value of the image (for 8-bit images, it is 255), and MSE is the Mean Squared Error, defined as:

\begin{equation}
\text{MSE} = \frac{1}{HW} \sum_{i=1}^{H} \sum_{j=1}^{W} \left[ I(i,j) - \hat{I}(i,j) \right]^2 \quad
\end{equation}

where \( I \) represents the real image, \( \hat{I} \) represents the generated image, and \( H \) and \( W \) are the height and width of the image. The higher the PSNR value, the smaller the pixel-level difference between the generated and real images, meaning the image is clearer and has less distortion.

SSIM is a metric that measures the structural similarity of images, considering statistical features such as brightness, contrast, and structure. Its definition is:

\begin{equation}
\text{SSIM}(x, y) = \frac{(2\mu_x \mu_y + C_1)(2\sigma_{xy} + C_2)}{(\mu_x^2 + \mu_y^2 + C_1)(\sigma_x^2 + \sigma_y^2 + C_2)} \quad
\end{equation}

where \( \mu_x \) and \( \mu_y \) are the means of images \( x \) and \( y \), \( \sigma_x^2 \) and \( \sigma_y^2 \) are the variances, \( \sigma_{xy} \) is the covariance, and \( C_1 \) and \( C_2 \) are constants for stability. The SSIM value ranges from [0,1], with values closer to 1 indicating that the generated image is structurally more similar to the real image, aligning with human sensitivity to image structure.

LPIPS is a deep learning-based perceptual image similarity metric that compares the feature representations of two images in a deep neural network (such as AlexNet or VGG), measuring their differences in the perceptual space. The calculation can be briefly expressed as:

\begin{equation}
\begin{split}
\text{LPIPS}(x, y) = & \sum_l \frac{1}{H_l W_l} \sum_{h,w} \left\| w_l \odot \left( f_l^x(h, w) - f_l^y(h, w) \right) \right\|_2^2 \\
& - \left\| f_l^y(h, w) \right\|_2^2 \quad
\end{split}
\end{equation}

Here, $f_l^x$ and $f_l^y$ are the $l$-th-layer feature maps of images $x$ and $y$, weighted by $w_l$ and normalized over $H_l \times W_l$. LPIPS, unlike PSNR or SSIM, aligns with human perception and captures semantic differences; the smaller the score, the more the generated image resembles the real one.

\subsection{Comparisons with State-of-the-Art Methods}
To comprehensively evaluate AnatoMaskGAN, we compared it with mainstream generative models on two public datasets (Table \ref{tab:compare-results}). On the L2R-OASIS brain-MRI set it achieved 26.50 dB PSNR, 0.9229 SSIM, and an LPIPS of 0.0599—0.0477 lower than the SPADE baseline—demonstrating superior fidelity and structural accuracy. Even on the more challenging L2R-Abdomen CT dataset, with limited samples and high anatomical variability, it still obtained 21.98 dB PSNR, 0.8602 SSIM, and 0.0807 LPIPS, confirming strong few-shot generalization.

Qualitative results. AS shown in Fig. \ref{fig5} and Fig. \ref{fig6}, AnatoMaskGAN faithfully reconstructs fine anatomy while suppressing acquisition artifacts in brain MRI, whereas competing models exhibit blurred textures and missing boundaries. The pattern repeats for abdomen CT: our method preserves crisp organ edges and intact shapes, while baselines suffer from texture blur and broken boundaries. Because these gains stem from its structural modeling and spatial-feature fusion rather than data volume, AnatoMaskGAN delivers high-quality, clinically useful images even with scarce training data, underscoring its broad applicability in medical imaging.

\subsection{Ablation Studies }
We conducted a systematic ablation study on the L2R-OASIS and L2R-Abdomen CT datasets (Tables \ref{tab:ablation}), starting from a SPADE backbone and successively adding GNN-SASE, 3D Spatial-Noise, and G-TC to gauge each component’s impact on image quality, structural fidelity, and perceptual realism.

On L2R-OASIS, GNN-SASE immediately boosted PSNR by 1.72 dB and SSIM by 0.244 while sharply reducing LPIPS—demonstrating that its non-local cross-slice modeling faithfully recovers cortical folds and other fine structures—then 3D Spatial-Noise diversified features, curbed mode collapse, and stabilized training without sacrificing spatial coherence, and finally G-TC refined gray-level distributions and textures, driving LPIPS to its lowest point and nudging PSNR and SSIM even higher, particularly at the white-matter/gray-matter interface. On L2R-Abdomen CT, although numeric gains are subtler, the same three-stage pattern persists: GNN-SASE enforces boundary consistency, 3D Spatial-Noise enriches detail and lowers LPIPS, and G-TC secures the best overall PSNR and SSIM under anatomically varied, limited data—together endowing AnatoMaskGAN with exceptional structural fidelity, visual authenticity, and cross-dataset robustness.

\section*{Conclusion}
AnatoMaskGAN tackles the spatial incoherence and loss of detail in semantic-mask–to-image synthesis. A graph-based slice-feature fusion module encodes inter-slice dependencies, while 3-D noise injection and a combined gray-level/texture constraint give a near-3-D anatomical view without sacrificing 2-D efficiency. On the L2R-OASIS and L2R-Abdomen CT datasets, it lifts PSNR to 26.50 dB and SSIM to 0.8602, outperforming all baselines; ablations show that slice fusion, 3-D noise, and the texture classifier each add complementary gains across PSNR, SSIM, and LPIPS.

Remaining hurdles include maintaining cross-modal stability, trimming the graph-reasoning overhead, and coping with very small datasets. Future work will explore lighter slice-relationship encoders, cross-modal alignment, and few-shot or self-supervised pre-training to cut resource demands and extend use to rare-disease and privacy-restricted settings. By uniting spatial slice fusion with adaptive texture-gray modeling, AnatoMaskGAN can drive data augmentation, virtual contrast generation, model pre-training, and clinical training, promising wide medical impact.

\bibliographystyle{IEEEtran}
\bibliography{reference}
\end{document}